\documentclass[
 aip,
jcp,
 amsmath,amssymb,
 reprint,%
]{revtex4-1}

\usepackage{color}
\usepackage{subfigure}
\usepackage{graphicx}
\usepackage{amssymb}
\usepackage{bm}
\usepackage{hyperref}

\definecolor{mma1}{rgb}{0.3725,0.5098,0.7020}
\definecolor{mma2}{rgb}{0.8745,0.6078,0.2039}
\definecolor{mma3}{rgb}{0.507813,0.714844,0.2039}
\definecolor{mma4}{rgb}{0.9137,0.3882,0.2398}
\definecolor{mma5}{rgb}{0.529412, 0.47451, 0.694118}
\definecolor{mma6}{rgb}{0.764706, 0.431373, 0.152941}
\definecolor{mma7}{rgb}{0.376471, 0.619608, 0.776471}

\begin{document}

%

\def\hslash{\hbar}
\def\imag{i}
\def\grad{\vec{\nabla}}
\def\div{\vec{\nabla}\cdot}
\def\curl{\vec{\nabla}\times}
\def\DDt{\frac{d}{dt}}
\def\ddt{\frac{\partial}{\partial t}}
\def\ddx{\frac{\partial}{\partial x}}
\def\ddy{\frac{\partial}{\partial y}}
\def\lap{\nabla^{2}}
\def\divv{\vec{\nabla}\cdot\vec{v}}
\def\gradS{\vec{\nabla}S}
\def\vvec{\vec{v}}
\def\wc{\omega_{c}}
\def\<{\langle}
\def\>{\rangle}
\def\Tr{{\rm Tr}}
\def\Csch{{\rm csch}}
\def\Coth{{\rm coth}}
\def\Tanh{{\rm tanh}}
\def\g2{g^{(2)}}
\newcommand{\al}{\alpha}
\newcommand{\la}{\lambda}
\newcommand{\del}{\delta}
\newcommand{\om}{\omega}
\newcommand{\ep}{\epsilon}
\newcommand{\pd}{\partial}
\newcommand{\bra}{\langle}
\newcommand{\ket}{\rangle}
\newcommand{\bbra}{\langle \langle}
\newcommand{\kket}{\rangle \rangle}
\newcommand{\non}{\nonumber}
\newcommand{\be}{\begin{equation}}
\newcommand{\ee}{\end{equation}}
\newcommand{\bea}{\begin{eqnarray}}
\newcommand{\eea}{\end{eqnarray}}

\title{On the derivation of exact eigenstates of the generalized squeezing operator}

\author{Andrey Pereverzev}
\email[email: ]{pereverzeva@missouri.edu}
\affiliation{Department of Chemistry,
University of Missouri\\
 Columbia, MO 65211}

\author{Eric R. Bittner}
\email[email: ]{ebittner@central.uh.edu}
\affiliation{Department of Chemistry,
University of Houston \\ Houston, TX 77204}

\date{\today}

\begin{abstract}
We construct the states that are invariant under the action of the generalized squeezing
operator $\exp{(z{a^{\dagger k}}-z^*a^k)}$ for arbitrary positive  integer $k$. The states are given explicitly in the number representation. We find that for a 
given value of $k$ there are $k$ such states. We show that the states behave as $n^{-k/4}$ when occupation number $n\to\infty$. This implies  
that for any $k\geq3$ the states are normalizable. For a given $k$, the expectation values of
operators of the form $(a^{\dagger} a)^j$ are finite for positive integer $j < (k/2-1)$ but diverge for integer $j\geq (k/2-1)$.
For $k=3$ we also give an explicit form of these states in the momentum representation in terms of Bessel functions.
\end{abstract}


\maketitle
\section{Introduction} 
The concept of squeezing  plays one of the central roles in quantum optics. 
Squeezed states 
facilitate
measurement and communication in a way not possible 
with the coherent states which are produced from quantum  
vacuum.  
Squeezed states are characterized by the phase-space distribution 
of the associated momentum-like ($\hat P$) and position-like ($\hat X$ )
quadrature variables 
of the field.   Their variances obey the Heisenberg principle
$\Delta \hat X \Delta \hat P \ge 1/4$.   Vacuum, coherent, and squeezed states 
minimally satisfy this inequality and a coherent state is realized when
$\Delta \hat X  = \Delta \hat P $.  A squeezed state is produced when 
either of the quadratures is increased at the expense of the other. 
Under purely harmonic time evolution, squeezed states remain squeezed  and, therefore,
always minimally satify the Heisenberg relation. However, they will evolve into non-squeezed states if non-harmonic perturbations are introduced to the Hamiltonian.
Refs.\citenum{Dodo} and \citenum{Illu} provide  extensive lists of references that deal with various aspects of squeezing. 

The mathematical realization of a squeezed state in the simplest case is given in terms of the squeezing operator
$U_2(z)=\exp{((z{a^{\dagger 2}}-{z^*a^2})/2)}$ acting on the vacuum state. Here  $a^{\dagger}$ and $a$ are creatrion and annihilation operators and $z$ is a compex-valued parameter. 
Over time, attempts to
generalize this operator to include higher order processes have been made. Different types of  generalizations have been investigated. 
Some of these generalizations involve exponentials of operators that are elements of closed algebras.
\cite{Lo1,March,Zelaya,Raffa}
By contrast, in this work we consider the generalization of the squeezing operator 
of the form 
\be
U_k(z)=\exp{(z{a^{\dagger k}}-{z^*a^k})}
\ee
with integer $k \geq 3$. This kind of generalization turned out to be quite nontrivial.
On one hand, it was shown by Fisher et al. \cite{Fisher} that the vacuum
to vacuum probability amplitude $\bra 0|U_k(z)|0\ket$ has a zero radius of convergence as a power series with respect to $z$, for  $k > 2$. On the other hand, Braunstein
 and McLachlan demonstrated numerically \cite{Braunstein1,Braunstein2} that such expressions can still be well defined. Some properties of operator $z{a^{\dagger k}}-{z^*a^k}$ were discussed by Nagel.\cite{Nagel} The exact eigenstates of this operator for the case of $k=2$ were also constructed by Lo.\cite{Lo2}
Elyutin and Klyshko
showed\cite{Elyutin} that the average occupation number of an arbitrary initial state acted upon by the
unitary operator $U_3(z)$ diverges to infinity for a {\it finite} $z$. All of these challenges lead to various attempts to modify the operators $a^{\dagger}$ and $a$ in $U_k(z)$ in such a way that no question of convergence would arise. \cite{D'Ari1,D'Ari2,Buz}

The $k$-squeezed states can be physically realized using the following argument. 
Consider the following unitary transformation of the harmonic Hamiltonian $H_0=\hbar\omega a^{\dagger}a$
\be
\tilde{H}=U_k(z)H_0U_k^{-1}(z) \label{ut} 
\ee
This expression can be evaluated explicitly only for $k=1$ and $2$.\cite{Wagner} However, for small $z$ we can expand the right-hand side of Eq. (\ref{ut}) in powers of $z$ and keep only the zeroth and first order terms. Application of the commutation rules for $a^{\dagger}$ and $a$ leads to
\be
\tilde{H}\approx H_0 -k(za^{\dagger k}+z^*a^k) \label{transformed} 
\ee
Thus, a $k$-th order squeezing would result 
from a $k$-th order non-linearity within an optically 
pumped system.

In this paper we construct $k$ exact degenerate eigenstates of operator $U_k(z)$ that have an eigenvalue equal to one and investigate some of their properties.

\section{Invariant states of operator $U_k(z)$ in the number representation} \label{first}
 Without the loss of generality we can limit ourselves to the case where $z$ is
real $z=z*=r$ and consider the eigenstates of the Hermitian operator
\be
M_k=ir({a^{\dagger k}}-{a}^k).
\ee

The eigenvalues of $M_k$ are not known except for the cases of $k=1, 2$. \cite{Nagel}
In general, the eigenvalue problem of this operator written in the number representation leads to a
 three term recurrence relation. However, in the case of zero eigenvalue
  (assuming that it exists) the recurrence relation involves only two terms and as a result
   explicit eigenstates can be obtained. It is clear that since
    these eigenstates of $M_k = ir( {a^{\dagger k}}-a^k$) have zero eigenvalues they are also eigenstates of $U_k(r)$
with eigenvalue equal to one, or, in other words, invariant under the action of $U_k(r)$.

Although we will be primarily interested in the case of $k\geq 3$, results of this section also apply to $k=1$ and $2$.
Note that  $U_k(z)$ commutes with operator  $G_k=\exp{(i\frac{2\pi}{k}a^{\dagger}a)}$. 

In the case of the zero eigenvalue 
the eigenvalue equation in the number representation reads
\be
\bra n|M_k|\psi_k^{\alpha}\ket=0,
\ee
where $\alpha$ is the degeneracy index.
Acting on $\bra n| $ from the right with the creation and annihilation operators in  $M_k$ we obtain the following recurrence relation
\be
\bra n+k|\psi_k^{\alpha}\ket=
\sqrt{\frac{n(n-1)(n-2)...(n-k+1)}{(n+1)(n+2)...(n+k)}}\bra n-k|\psi_k^{\alpha}\ket \label{recur}
\ee
and the following $k$ conditions 
\be
\bra k|\psi_k^{\alpha}\ket=0,\bra k+1|\psi_k^{\alpha}\ket=0,...\bra 2k-1|\psi_k^{\alpha}\ket=0.
\ee
By iteratively applying recurrence relation of Eq. (\ref{recur}) starting with $\bra 0|\psi_k^{\alpha}\ket$, 
$\bra 1|\psi_k^{\alpha}\ket$, ...$\bra k-1|\psi_k^{\alpha}\ket$,
we obtain after some algebra $k$ degenerate zero eigenstates of $M_k$. The non-vanishing components of these eigenstates have the following form
\bea
\bra \alpha+2mk |\psi_k^{\alpha}\ket&=&\frac{1}{\sqrt{c(k,\alpha)}}\frac{{(2k)}^{km}}{\sqrt{(\alpha+2mk)!}}  \non \\
& &\times \prod_{i=1}^k\Gamma\left(m+\frac{\alpha+i}{2k}\right). \label{functions}
\eea
Here integer $\alpha$ is the degeneracy index that can take values from $0$ to $(k-1)$ and integer $m$ runs from $0$ to infinity. 
The occupation number is given by $n = \alpha + 2mk$. Number states with occupation numbers
that do not satisfy the last equation do not contribute to the eigensates of $M_k$.
$\Gamma(x)$ denotes the gamma function and $c(k,\alpha)$ is the normalization constant given by 
\begin{widetext}
\be
c(k,\alpha)=_{k+1}\!\!F_k\left(1,\frac{1+\alpha}{2k},\frac{2+\alpha}{2k},...,\frac{k+\alpha}{2k};\frac{k+1+\alpha}{2k},\frac{k+2+\alpha}{2k},...,\frac{2k+\alpha}{2k};1\right)\frac{\prod_{i=1}^k\left(\Gamma(\frac{i+\alpha}{2k})\right)^2}{\alpha!}, \label{norm}
\ee 
\end{widetext}
where
$_pF_q(x_1,x_2,...,x_p;y_1,y_2,...,y_q;z)$ is the generalized hypergeometric series.
The eigenfunctions given by Eq. (\ref{functions}) are monotonically decreasing functions of $m$.

\begin{figure*}
 \subfigure[]{\includegraphics[width=0.6\columnwidth]{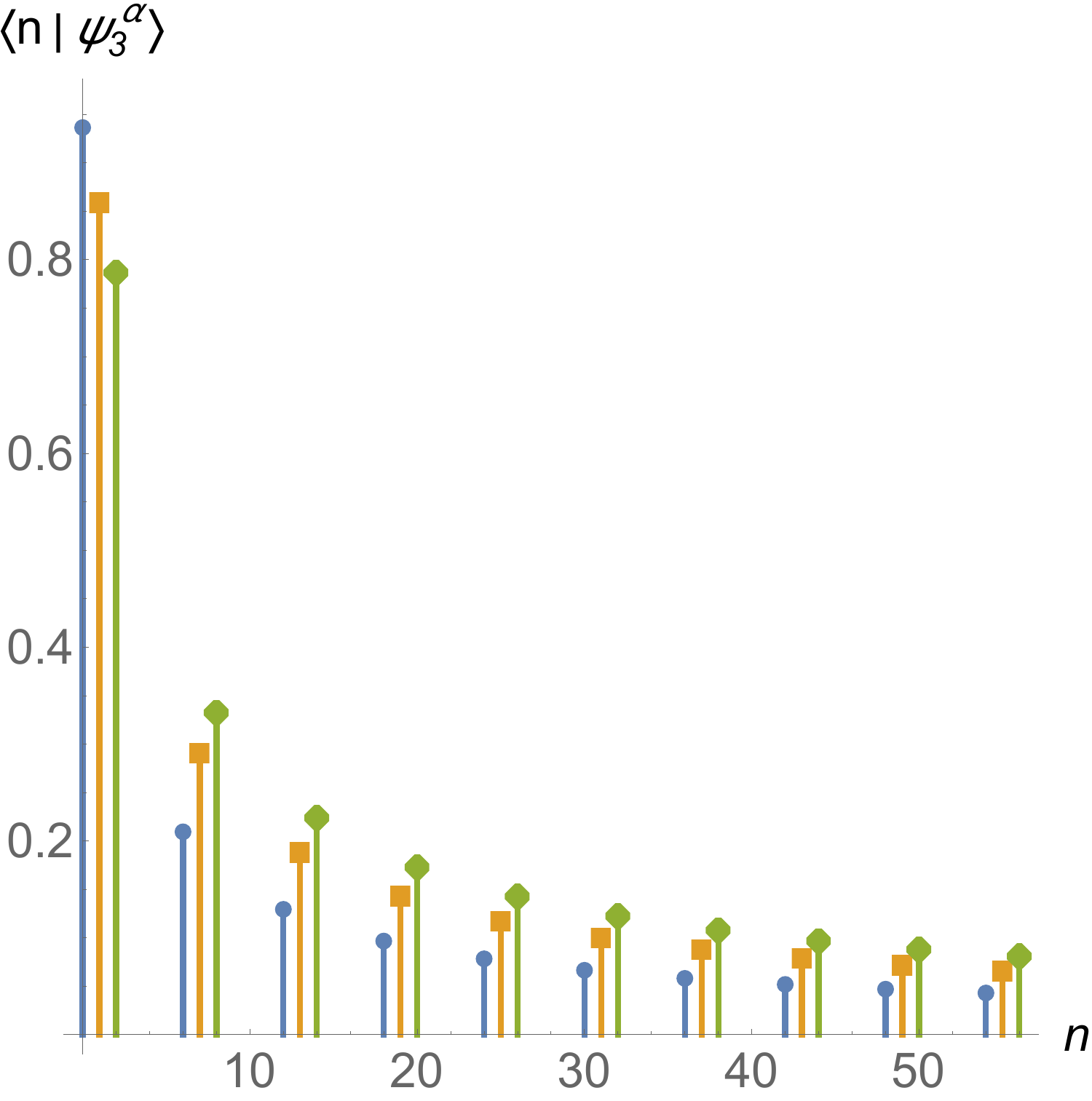} }
 \subfigure[]{\includegraphics[width=0.6\columnwidth]{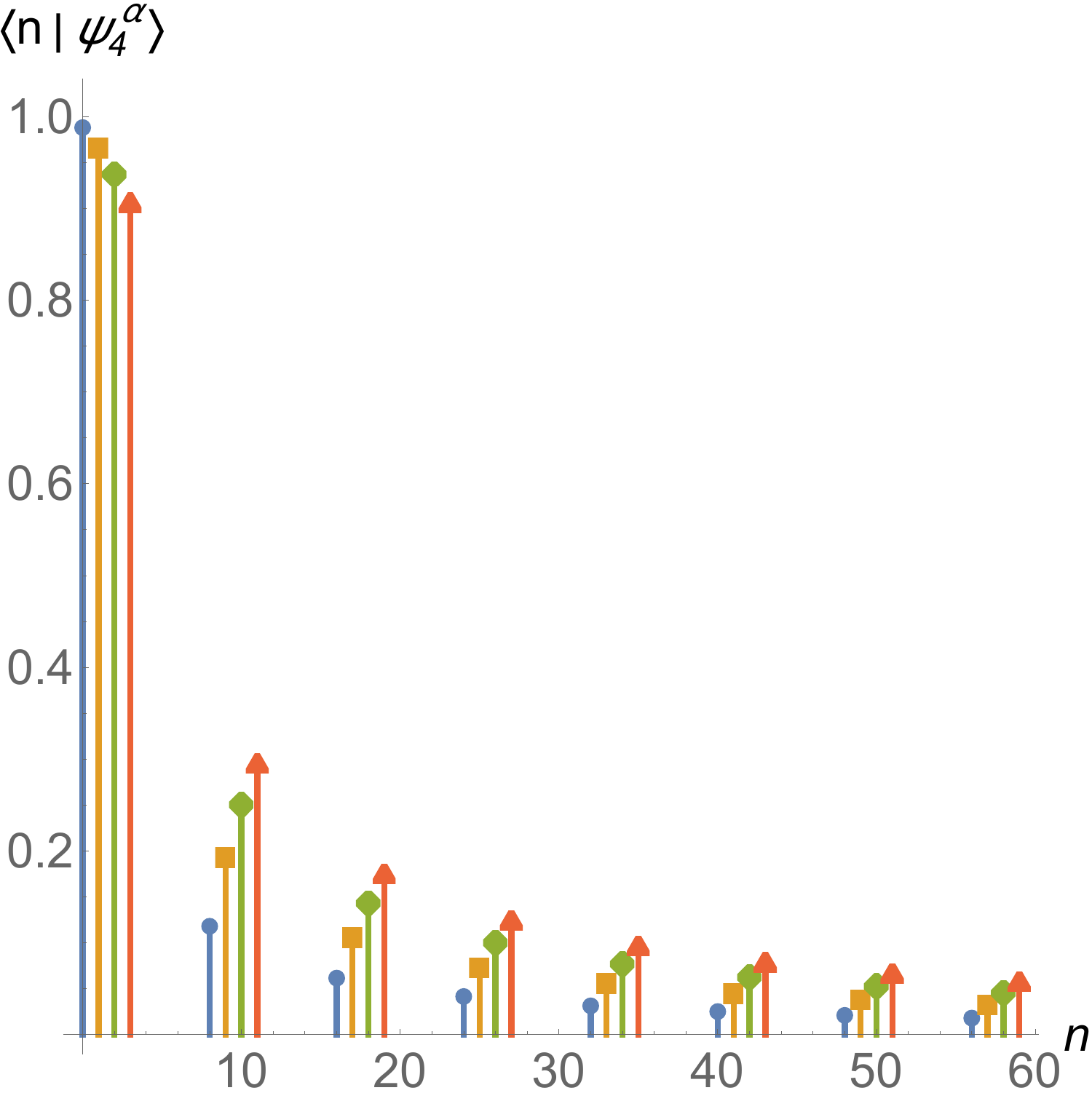} }\\
 \subfigure[]{\includegraphics[width=0.6\columnwidth]{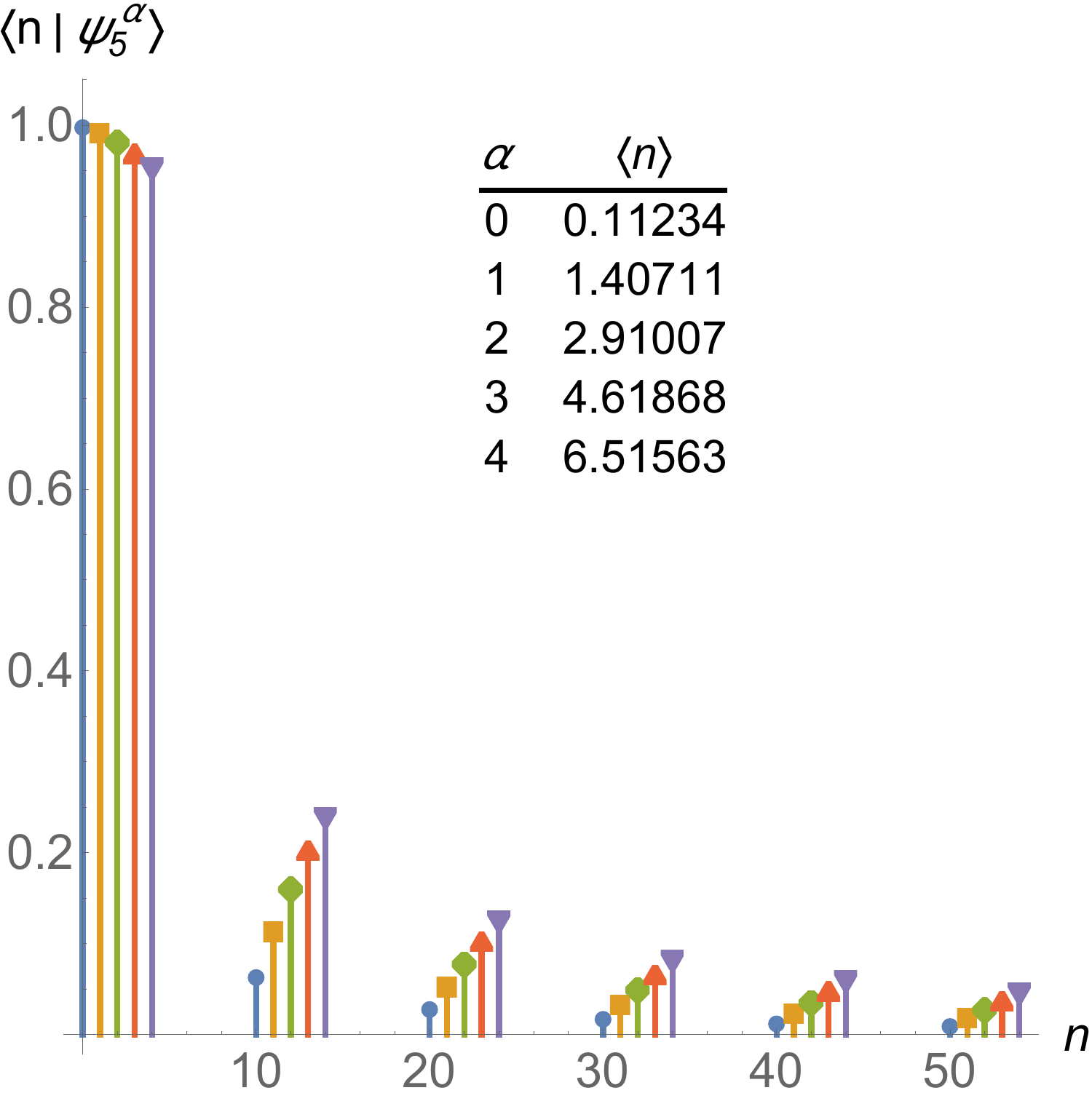}}
 \subfigure[]{\includegraphics[width=0.6\columnwidth]{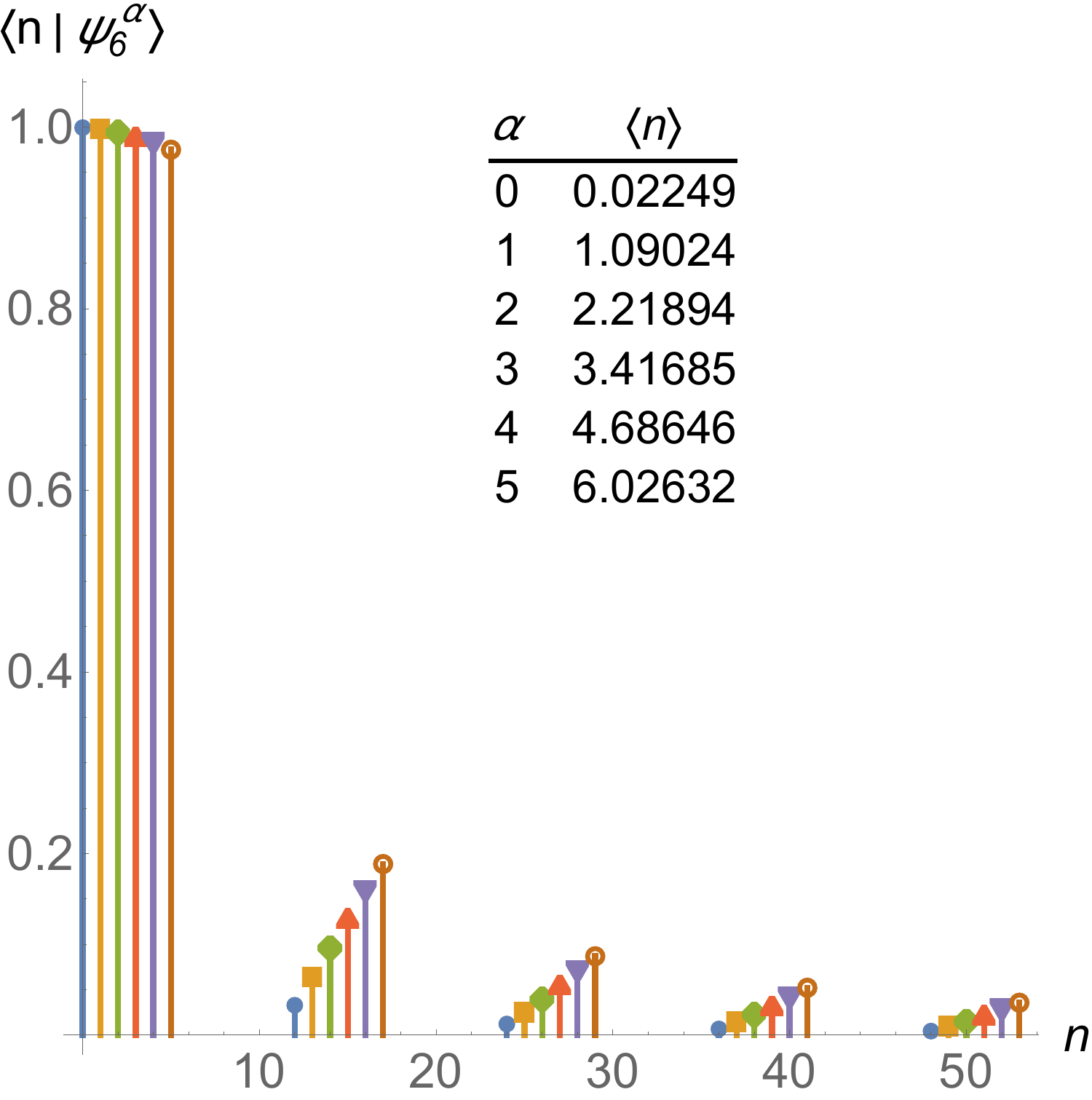} }
 \subfigure[]{\includegraphics[width=0.6\columnwidth]{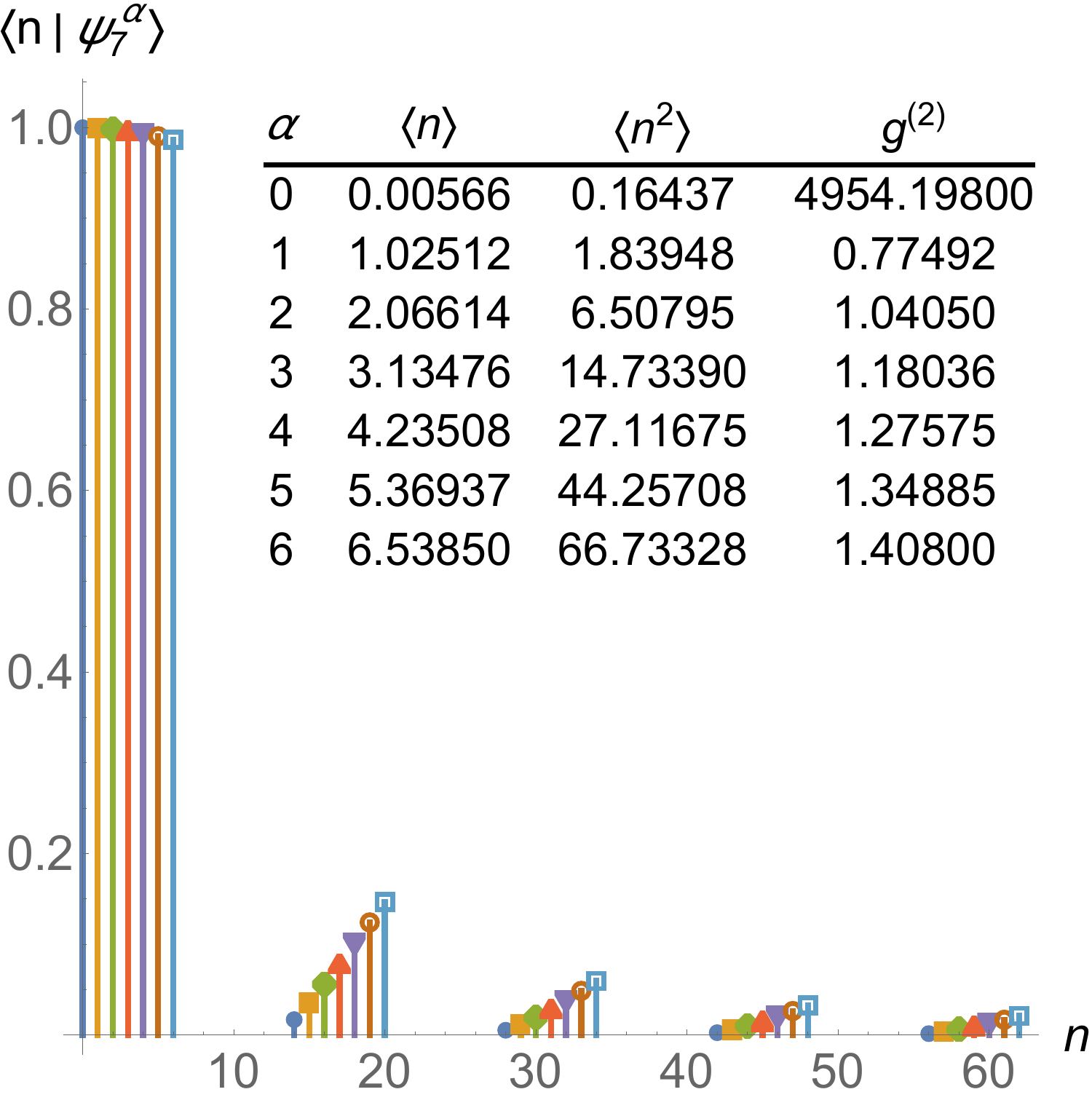} }
  
 \caption{\label{nrepr}Eigensates $|\psi_k^\alpha\rangle$ in the number represenation for $k = 3$ through $7$ and permitted values of $\alpha$. Zero-value components of the eigenstates are not shown.
Inset tables give the numerical expectation 
values of $\langle n\rangle$, $\langle n^2\rangle$, and $g^{(2)}$ for the cases when they are finite. 
(
$\textcolor{mma1}{\bullet}: ~\alpha = 0$, 
$\textcolor{mma2} {\blacksquare}:~\alpha = 1 $,
$\textcolor{mma3} { \blacklozenge}:~\alpha = 2 $,
$\textcolor{mma4} { \blacktriangle}:~\alpha = 3 $,
$\textcolor{mma5} { \blacktriangledown}:~\alpha = 4 $,
$\textcolor{mma6} { \circ}:~\alpha = 5 $,
$\textcolor{mma7} { \square}:~\alpha = 6 $
)
 }
 \label{nrepr}
 \end{figure*}

 Note that each of the $|\psi_k^\alpha\ket$ is also an eigenstates of operator $G_k$ with eigenvalue 
$\exp{(i\frac{2\pi}{k}\alpha)}$. The asymptotic 
behavior of functions (\ref{functions}) for large $m$ can be obtained 
with the help of  Stirling's expansions for the gamma functions 
and the factorial. After some algebra we obtain
for large $m$
\be
\bra \alpha+2mk |\psi\ket \sim d(k,\alpha)m^{-k/4}.
\ee
Here the prefactor $d(k,\alpha)$ is given by
\be
d(k,\alpha)=\frac{1}{\sqrt{c(k,\alpha)}}\frac{{(2\pi)}^{({2k-1})/{4}}}{{(2k)}^{{(2\alpha+1)}/{4}}}.
\ee
Since the square of the eigenstate behaves as $m^{-k/2}$ for large $m$ we can conclude that the
norm is finite for any $k\geq 3$. This is because the series of the form 
$\sum_{m=1}^\infty m^{-p}$ converges when $p>1$ and diverges when $p\leq 1$.  For $k=1$ and $2$ the norm (see Eq.(\ref{norm})) diverges in agreement with 
the known exact results for these cases. \cite{Lo2,Nagel} Similarly, we can see that the average for the number operator $\hat n=a^{\dagger}a$ is divergent for $k<5$,
the average $\hat n^2$ diverges for $k<7$, etc. In general, for a given $k$ the expectation values of operators of the form ${\hat n}^j$ diverge for the integer $j\geq (k/2-1)$. 

If we define dimensionless coordinate and momentum operators $\hat X=\sqrt{\frac{1}{2}}(a^{\dagger}+a)$, $\hat P=i\sqrt{\frac{1}{2}}(a^{\dagger}-a)$ then their expectation values for states given by Eq. (\ref{functions}) vanish. However, if superpositions of the degenerate states (\ref{functions}) are considered then, in
general, the average  of $\hat X$ and $\hat P$ will diverge for $k\leq3$ but converge
for $k \geq 4$.
Expectation values of $\hat X^2$ and $\hat P^2$ behave in the same way as that for $a^{\dagger}a$, namely, diverge for 
$k=3$ and $k=4$, but remain finite for $k\geq 5$.
The divergence of the expectation value of the number operator for $k=3$ and $k=4$ implies infinite average energy for these states. 

Fig. \ref{nrepr} shows $\bra n|\psi_k^\alpha\ket$'s as functions of occupation number $n$ for $k=3$ through $k=7$.  The inset tables for $k=5$ through $k=7$
  give the computed $\langle \hat n \rangle$,  $\langle \hat n^2 \rangle$, and second-order 
 intensity correlator  $g^{(2)} =\langle (a^\dagger)^2 a^2\rangle/\langle \hat n\rangle^2=(\langle \hat n^2 \rangle-\langle \hat n\rangle)/\langle \hat n\rangle^2$,
 for each allowed value of the degeneracy index $\alpha$.   The $g^{(2)}$ correlator is a particularly
 useful quantity since it gives the probability of detecting two simultaneous photons normalized 
 by the probability of detecting two photons from a random source.   One can generalize this to 
 $g^{(k)} = \langle (a^\dagger)^k a^k\rangle/\langle \hat n \rangle^k$; however, such terms 
 will diverge for reasons given above.

\section{Invariant states of $U_3(z)$ in the momentum representation}
\begin{figure}
 \includegraphics[width=\columnwidth]{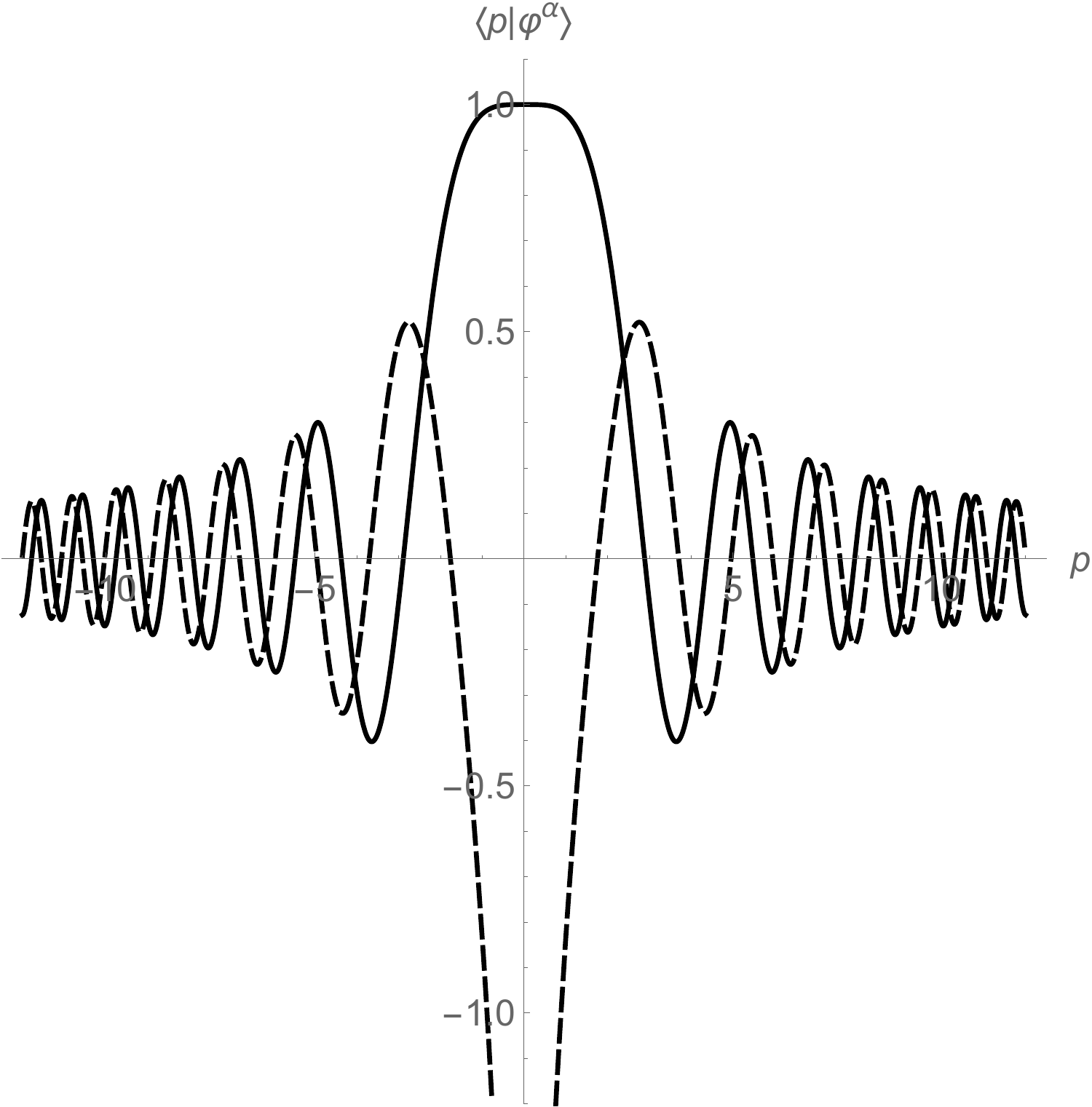} 
 \caption{\label{bessel} Functions $\bra p|\varphi^1\ket = J_0\left(\frac{p^2}{2\sqrt3}\right)$ (solid curve) and $\bra p|\varphi^2\ket = Y_0\left(\frac{p^2}{2\sqrt3}\right)$ (dashed curve).}
 \end{figure}

\begin{figure}
 \includegraphics[width=\columnwidth]{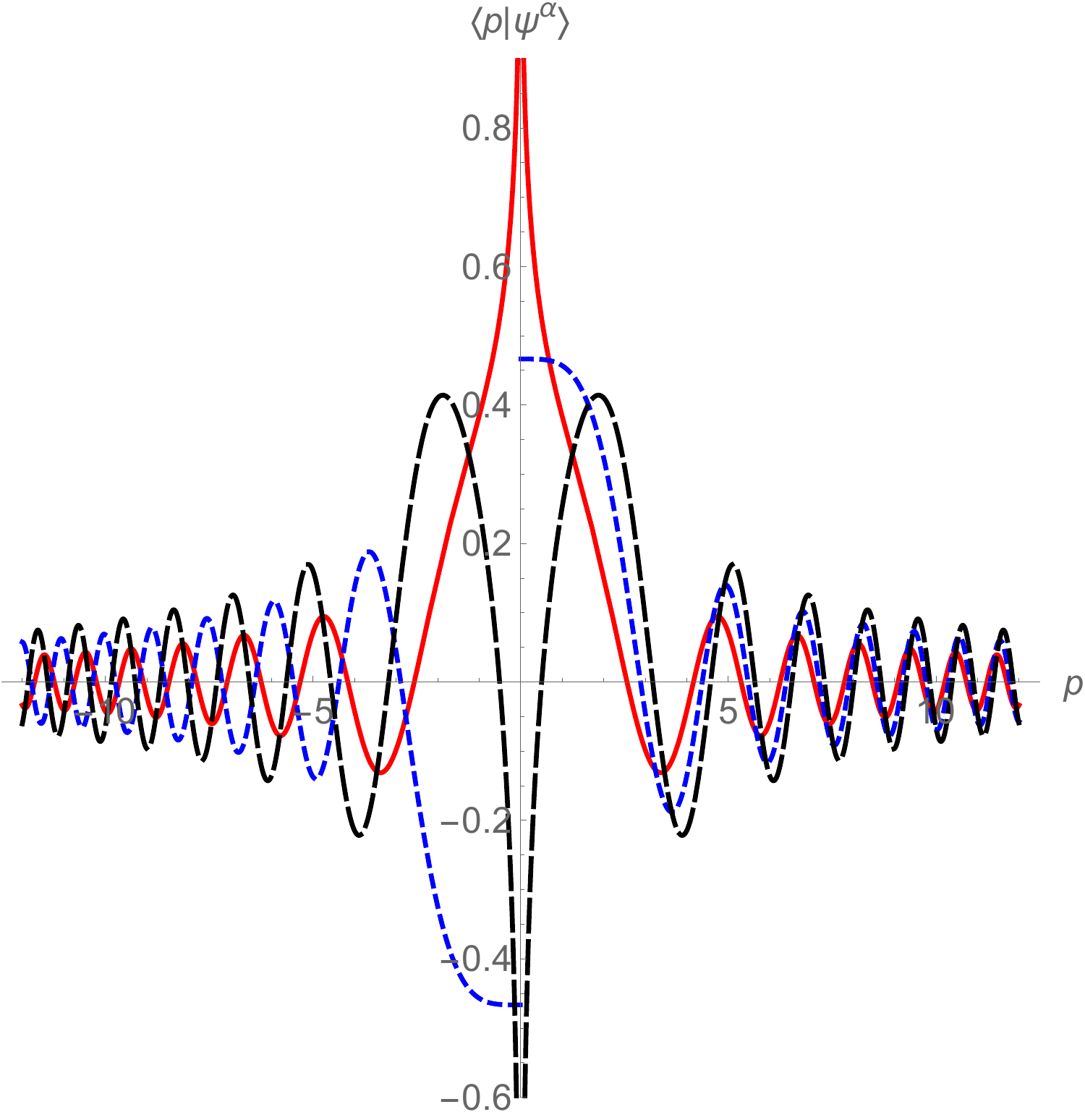} 
 \caption{\label{psi2s} Functions $\bra p|\psi^0\ket$ (solid red), $\bra p|\psi^1\ket$ (blue, short dashes), and $\bra p|\psi^2\ket$ (black, long dashes).}
 \end{figure}

Since eigenstates $\bra n |\psi_k^\alpha\ket$ decay slowly as functions of $n$,
it is of interest to consider their behavior in a continuum basis, such as coordinate or momentum representations. In this section we will costruct the invariant states of $U_k(z)$ in momentum representation for 
$k=3$. We chose momentum over coordinate representation to demonstrate an interesting mathematical point that will be mentioned below.
Rewriting  $a$ and $a^{\dagger}$ in terms of dimensionless coordinate and momentum operators as
$a=\sqrt{\frac{1}{2}} (\hat X+i \hat P)$ and $a^{\dagger}=\sqrt{\frac{1}{2}} (\hat X-i \hat P)$, inserting them into $M_3$, and using momentum representation we obtain the following eigenvalue equation for the zero eigenvalue
\be
\left(p\frac{d^2}{dp^2}+\frac{d}{dp}
+\frac{1}{3}p^3\right)\bra p|\psi^{\alpha}\ket=0. \label{second}
\ee
Here we suppress subscript $3$ for $k=3$ in the wave function to simplify the notation. This is a second order ordinary differential equation and its two independent solutions are given by \cite{Zai}
\bea
\bra p|\varphi^1\ket&=&J_0\left(\frac{p^2}{2\sqrt3}\right),\nonumber \\  
\bra p|\varphi^2\ket&=&Y_0\left(\frac{p^2}{2\sqrt3}\right),
\eea
where $J_{0}(x)$ and $Y_{0}(x)$ are the zeroth order Bessel functions of the first and second kind, respectively. Note that functions $\bra p |\varphi^1\ket$ and $\bra p |\varphi^2\ket$ are neither orthogonal to each other nor normalized. Plots of these functions are shown in Fig. \ref{bessel}.
The behavior of $\bra p|\varphi^1\ket$ and $\bra p|\varphi^1\ket$ for large $p$ is determined by the asymptotic behavior of the Bessel functions
\bea
\bra p|\varphi^1\ket 
&\sim& \frac{2(3^{1/4})}{\sqrt{\pi}p}\cos\left(\frac{p^2}{2\sqrt3}-\frac{\pi}{4}\right),\label{Asy1}\\
\bra p|\varphi^2\ket 
&\sim& \frac{2(3^{1/4})}{\sqrt{\pi}p}\sin\left(\frac{p^2}{2\sqrt3}-\frac{\pi}{4}\right). \label{Asy2}
\eea
Function $\bra p|\varphi^2\ket$ has a logarithmic singularity at $p=0$.
Both $\bra p |\varphi^1\ket$ and $\bra p |\varphi^2\ket$ are even functions of $p$ and, therefore must be linear combinations of eigenstates $\bra p |\psi^0\ket$ and $\bra p |\psi^2\ket$. 
The obvious question then is what happened to the third eigenfunction $|\psi^1\ket$ which must be odd in the momentum representation. The answer to this question comes from noting that Eq. (\ref{second})
is singular at $p=0$. This becomes obvious once both sides of Eq. (\ref{second}) are divided over by $p$ to bring the equation to the standard form. The third solution is obtained by reflecting 
$\bra p |\varphi^1\ket$ taken from $-\infty$ to $0$ with respect to the $p$ axis. Thus,  
\be
\bra p |\varphi^3\ket= J_0\left(\frac{p^2}{2\sqrt3}\right)\mbox{sgn}(p), \label{phi}
\ee
where $\mbox{sgn}(p)$ is the sign function. It is easy to verify that $\bra p |\varphi^3\ket$ is indeed a solution of Eq. (\ref{second}) since 
the differentiations at the vicinity of the ``step" at $0$ give zero contribution. 
When properly normalized (up to an arbitrary phase factor), eigenfunctions $\bra p|\psi^{\alpha}\ket$'s for the case of $k=3$ in Eq. (\ref{second}) 
are expressed through $\bra p |\varphi^1\ket$, $\bra p |\varphi^2\ket$, and $\bra p |\varphi^3\ket$ as follows
\bea
\bra p|\psi^0\ket&=&a_0\left(\bra p|\varphi^1\ket-\frac{1}{\sqrt3}\bra p|\varphi^2\ket\right), \label{psi0}\\
\bra p|\psi^1\ket&=&a_1\bra p|\varphi^3\ket, \label{psi1}\\
\bra p|\psi^2\ket&=&a_2\left(\bra p|\varphi^1\ket+\frac{1}{\sqrt3}\bra p|\varphi^2\ket\label{psi2}\right),   
\eea
where coefficients $a_0$,  $a_1$, and $a_2$ are given by
\bea
a_0 &=&\left(\frac{(2\sqrt3-3)\pi}{4}\right)^\frac{1}{4}\frac{\Gamma(\frac{3}{4})}{\Gamma(\frac{1}{4})}, \\
a_1 &=&\left(\frac{2\pi}{\sqrt3}\right)^\frac{1}{4}\frac{\Gamma(\frac{3}{4})}{\Gamma(\frac{1}{4})}, \\
a_2 &=&\left(\frac{(2\sqrt3+3)\pi}{4}\right)^\frac{1}{4}\frac{\Gamma(\frac{3}{4})}{\Gamma(\frac{1}{4})}.
\eea
Functions $\bra p|\psi^\alpha\ket$ are shown in Fig. \ref{psi2s}.
Their asymptotic behavior is determined by the asymptotics of $\bra p|\varphi^1\ket$ and $\bra p|\varphi^2\ket$ given by Eqs. (\ref{Asy1}) and (\ref{Asy2}). All $\bra p|\psi^\alpha\ket$'s show slow oscillating decay 
for large $|p|$, $\bra p|\psi^0\ket$ and $\bra p|\psi^2\ket$ have logarithmic singularities at $p=0$. 
All three functions, however, are square integrable in agreement with the results of Sec. \ref{first}.
Explicit solutions can also be obtained in the coordinate representation either by Fourier transforming $\bra p|\psi^\alpha\ket$'s or by solving Eq. (\ref{second}) rewritten in the coordinate representation. We will not consider them in this paper. Note, however, that in the coordinate representation Eq. (\ref{second}) is the third order differential equation and, therefore, the issue of the "missing solution" does not arise. 

Finally, let us note the following interesting property of the eigenvalue equation for $M_3$ in the  momentum representation - it is solvable in the Sturm-Liouville sense, namely, if we define
functions
\bea
f^+_l(p)&=&J_0\left(\frac{l p^2}{2\sqrt3}\right)\theta(p)\theta(l), \non \\ 
f^-_l(p)&=&J_0\left(\frac{l p^2}{2\sqrt3}\right)\theta(-p)\theta(-l),
\eea
where $\theta(x)$ is the Heaviside step function,
it can be verified that
\be \left(p\frac{d^2}{dp^2}+\frac{d}{dp}
+\frac{1}{3}p^3\right)f^{\pm}_l(p)
=\frac{(1-l^2)}{3}p^3f^{\pm}_l(p).
\ee
Thus, $f^{\pm}_l(p)$ is an eigenfunction with eigenvalue $\frac{1}{3}(1-l^2)$ and weight
function $p^3$. Functions $f^{\pm}_l(p)$ form a complete set,
\be
\frac{1}{6}\int_{-\infty}^{\infty}\!dl\, l \big(f^{+}_l(p)f^{+}_l(p')+f^{-}_l(p)f^{-}_l(p')\big)
=\frac{1}{p^3}\delta(p-p').
\ee
It appears, however,  that this solution cannot be used to 
construct the spectrum of the exponential operator $U_3(z)$. 

\section{Discussion}
We explicitly constructed some of the eigenstates of the generalized squeezing operator $U_k(z)=\exp{(z{a^{\dagger k}}-{z^*a^k})}$ in the number representations and showed that they are normalizable for 
$k\geq3$ but have divergent expectation values for operators $(a^{\dagger} a)^j$  for the integer $j\geq (k/2-1)$. We obtained only $k$ eigenstates of $U_k(z)=\exp{(z{a^{\dagger k}}-{z^*a^k})}$. If we assume that the remaining eigenstates of $U_k(z)$ have similar convergence properties this would imply that operator $U_k(z)$ has a spectral resolution in the Hilbert space.
Moreover, the states that we found can become useful for approximate treatments of operator $U_k(z)$. In  particular, if these operators are approximated by finite dimensional matrices, then the suitable basis sets can be chosen to have convergence properties similar to the states that were considered in this paper.
 Due to their interesting  properties the states $|\psi_k^{\alpha}\ket$ can be of also interest for mathematical physics applications such as the theory of generalized squeezed and coherent states \cite{Dodo} or the theory of wavelets.\cite{Hoffman}

\begin{acknowledgments}
The work at the University of Houston was funded in
part by the  National Science Foundation (
CHE-1664971,   
CHE-1836080,  
DMR-1903785    
) and the Robert A. Welch Foundation (E-1337). 
ERB wishes to acknowledge Dr. ARS Kandada
for discussions and 
sharing his preliminary results of squeezed
states. 

\end{acknowledgments}

\section*{Data Availability}
The data that support the findings of this study are available from the corresponding author upon reasonable request.

\end{document}